\documentstyle[aps,prl,psfig,floats]{revtex}
\begin{document}
\draft

\twocolumn[\hsize\textwidth\columnwidth\hsize\csname @twocolumnfalse\endcsname

\title{Phase Separation in One-Dimensional Driven Diffusive Systems\\}

\author{M. R. Evans$^{(1)}$, Y. Kafri$^{(2)}$, H. M. Koduvely$^{(2)}$,
and D. Mukamel$^{(2)}$}

\address{$^{1}$ Department of Physics and Astronomy, University of
        Edinburgh, Mayfield Road, Edinburgh EH9 3JZ, U.K.\\ $^{2}$
        Department of Physics of Complex Systems, The Weizmann Institute 
        of Science, Rehovot 76100, Israel\\[-2mm] $ $}
        
\date{\today}

\maketitle
\begin{abstract}
A driven diffusive model of three types of particles that exhibits
phase separation on a ring is introduced.  The dynamics is local and
comprises nearest neighbor exchanges that conserve each of the three
species.  For the case in which the three densities are equal, it is
shown that the model obeys detailed balance. The Hamiltonian governing
the steady state distribution in this case is given and is found to
have long range asymmetric interactions.  The partition sum and bounds
on some correlation functions are calculated analytically
demonstrating phase separation.
\end{abstract}
\vspace{2mm}
]

Driven diffusive systems have been a subject of extensive studies in
recent years \cite{Leb,dds}. Being driven by an external field, these
systems are governed by dynamics which generically does not obey
detailed balance, leading to steady states with non-vanishing
currents.  This class of systems provides a relatively simple
framework within which collective phenomena far from thermal
equilibrium may be studied \cite{HHZ,evans1}.\

Theoretical studies of models of driven diffusive systems reveal some
basic differences between phase transitions taking place under
equilibrium and non-equilibrium conditions.  For example it is well
known that phase transitions and spontaneous symmetry breaking are not
expected to take place in one dimensional ($1d$) systems in thermal
equilibrium at finite temperatures, as long as the interaction in the
system is short range. Recently it has been demonstrated that this is
not the case in systems far from thermal equilibrium \cite{Gacs}.
Indeed, a simple example of an open system with non-conserved order
parameter at the boundaries was shown to exhibit spontaneous symmetry
breaking \cite{evans1}.

A closely related problem is that of phase separation.  An interesting
question is whether $1d$ homogeneous systems (i.e. with no boundary
effects as in a ring geometry) are capable of exhibiting phase
separation, in cases where conserving dynamics is involved.  It has
been shown that inhomogeneities, such as defect sites or particles may
trigger the formation of macroscopic regions of high density bound to
the defect \cite{defect}.  Recent numerical studies of a model
associated with sedimentation of colloidal crystals have suggested
that phase separation may even occur in homogeneous non-equilibrium
systems \cite{lahiri}.

Phase separation is accompanied by coarsening phenomena in which the
typical domain size grows indefinitely with time \cite{Bray}.
Examples of equilibrium $1d$ systems with local dynamics which exhibit
coarsening are the zero temperature limit of kinetic Ising models and
the noiseless Landau-Ginzburg equation. However these systems do not
coarsen at finite temperature.

In the present Letter we introduce a simple model of phase separation
in a $1d$ driven diffusive system. This is a model of $3$ species of
particles on a ring in which nearest neighbors on the lattice are
exchanged with specific rates.  Thus the dynamics is local, fully
stochastic and conserves each of the three species. It is shown that
for the special case in which the average densities of the three
species are equal, the dynamics obeys detailed balance. In this case
the steady state distribution is shown to be given by a Hamiltonian
which has long range asymmetric interactions. The phase separation
which takes place in the model is explicitly demonstrated for this
case. It is argued that phase separation takes place in the general
case where the densities of the three species are unequal and that the
typical domain size coarsens as $\ln t$.  The model is easily
generalized to $n$ species and phase separation is found provided
$n>2$.\

{\it A. Definition of the model:}
The model is defined on a $1d$ lattice of length $N$ with periodic
boundary conditions. Each site is occupied by either an $A$, $B$, or
$C$ particle. The evolution is governed by random sequential dynamics
defined as follows: at each time step two neighboring sites are chosen
randomly and the particles of these sites are exchanged according to
the following rates
\begin{equation}
\label{eq:dynamics}
\begin{picture}(130,37)(0,2)
\unitlength=1.0pt
\put(36,6){$BC$}
\put(56,4) {$\longleftarrow$}
\put(62,0) {\footnotesize $1$}
\put(56,8) {$\longrightarrow$}
\put(62,13) {\footnotesize $q$}
\put(80,6){$CB$}
\put(36,28){$AB$}
\put(56,26) {$\longleftarrow$}
\put(62,22) {\footnotesize $1$}
\put(56,30) {$\longrightarrow$}
\put(62,35) {\footnotesize $q$}
\put(80,28){$BA$}
\put(36,-16){$CA$}
\put(56,-18) {$\longleftarrow$}
\put(62,-22) {\footnotesize $1$}
\put(56,-14) {$\longrightarrow$}
\put(62,-9) {\footnotesize $q$}
\put(80,-16){$AC$.}
\end{picture}
\end{equation}
\vspace{0.3cm}

\noindent The rates are cyclic in $A$, $B$ and $C$ and conserve the number of
particles of each type.\

Consider a system with $N_A$ particles of type $A$, $N_B$ of type $B$,
and $N_C$ of type $C$. For $q=1$ the particles undergo symmetric
diffusion and the system is disordered. However for $q \neq 1$ the
particle exchange rates are biased. Since the model is invariant, for
example, under the exchange $A \to B$ and $q \to 1/q$, it is
sufficient to consider $q<1$. In this case the bias drives, say, an
$A$ particle to move to the left inside a $B$ domain, and to the right
inside a $C$ domain. Therefore, starting with an arbitrary initial
configuration, the system reaches after a relatively short transient
time a state of the type $\ldots AABBCCAAAB \ldots$ in which $A,B$ and
$C$ domains are located to the right of $C$, $A$ and $B$ domains,
respectively.  Due to the bias $q$, the domain walls $\ldots AB
\ldots$, $\ldots BC \ldots$, and $\ldots CA \ldots$, are stable, and
configurations of this type are long lived. In fact, the domains in
these configurations diffuse into each other and coarsen on a time
scale of the order of $q^{-L}$, where $L$ is a typical domain size in
the system.  This leads to the growth of the typical domain size as $(
\ln t)/\vert\ln q \vert$. Eventually the system phase separates into
three domains of the different species of the form $A \ldots AB \ldots
BC \ldots C$. A finite system does not stay in such a state
indefinitely.  For example, the $A$ domain breaks up into smaller
domains in a time of order $q^{-min \lbrace N_B,N_C \rbrace}$. In the
thermodynamic limit, however, when the density of each type of
particle is non vanishing, the time scale for the break up of
extensive domains diverges and we expect the system to phase
separate. Generically the system supports particle currents in the
steady state. This can be seen by considering, say, the $A$ domain in
the phase separated state. The rates at which an $A$ particle
traverses a $B$ ($C$) domain to the right (left) is of the order of
$q^{N_B}$ ($q^{N_C}$). The net current is then of the order of
$q^{N_B}-q^{N_C}$, vanishing exponentially with $N$. This simple
argument suggests that for the special case $N_A=N_B=N_C$ the current
is zero for any system size.\

In the following, we show that the dynamics of the model satisfies
detailed balance for the special case $N_A=N_B=N_C$. The Hamiltonian
governing the steady state distribution is found to have long range
interactions. It is demonstrated that phase separation takes place in
the thermodynamic limit. The simple considerations given above
indicate that phase separation exists even when the number of
particles of the three species are not equal.

{\it B. Special case, $N_A = N_B =N_C$ :}
We specify a configuration by the set of numbers $\lbrace X_i \rbrace
= \lbrace A_i,B_i,C_i \rbrace$, where $A_i$, $B_i$ and $C_i$ are equal
to one if site $i$ is occupied by particle $A$, $B$ or $C$
respectively and zero otherwise.  We show that the dynamics satisfies
detailed balance with respect to the Hamiltonian ${\cal H}$ given by:
\begin{equation}
\label{eq: hamil}
{\cal H}(\lbrace X_i \rbrace)=\sum_{i=1}^{N-1}\sum_{j=i+1}^{N}
[C_iB_j-C_iA_j+B_iA_j].
\end{equation}   
So that, in the steady state
the probability of
finding the system in a configuration $\lbrace X_i \rbrace$ is given by:
\begin{equation}
\label{eq:weight}
W_N(\lbrace X_i \rbrace)=Z_N^{-1}q^{{\cal H}(\lbrace X_i \rbrace)},
\end{equation}
where the partition sum is given by $Z_N=\sum_{\lbrace
X_i\rbrace}q^{{\cal H}(\lbrace X_i \rbrace)}$ and the sum runs over
all states in which $N_A=N_B=N_C$.  Although expression (\ref{eq:
hamil}) for the Hamiltonian is not manifestly translationally
invariant, a form that is clearly invariant under translations can be
simply obtained from (\ref{eq: hamil}) but is less convenient for our
purposes. \

Before proving Eqs.(\ref{eq: hamil},\ref{eq:weight}) let us make a few
comments.  The $N$-fold degenerate zero energy ground states of the
Hamiltonian (\ref{eq: hamil}) comprise the fully separated
configuration $A\ldots AB \ldots BC\ldots C$ and any translation of
this configuration. Starting from a ground state, any other
configuration may be obtained by successive permutations of nearest
neighbor particles. The energy of such configuration may be calculated
by noting that any particle exchange against the bias costs one unit
of energy while an exchange in the direction of the bias results in a
gain of one unit.  The maximal energy of this Hamiltonian is $N^2/9$
and corresponds to a fully separated state $A \ldots AC \ldots CB
\ldots B$. It is of interest to note that although the dynamics
defined in (\ref{eq:dynamics}) is local, the resulting Hamiltonian
(\ref{eq: hamil}) is long range, in which each particle interacts with
all other particles. Also the interactions are asymmetric in the sense
that the Hamiltonian is not invariant under space inversion.  The
temperature of the system is given by $T=1/\vert \ln q \vert $. \

To prove Eqs.(\ref{eq: hamil},\ref{eq:weight}) one can check that
$W_N(\lbrace X_i \rbrace)$ satisfies detailed balance with respect to
the dynamics (\ref{eq:dynamics}). For example it is easy to see that
an exchange of particles in the bulk of the lattice $\ldots AB \ldots
\rightarrow \ldots BA \ldots$ satisfies
$qW_N(...AB...)=W_N(...BA...)$. Similarly, particle exchange between
site $1$ and $N$, say, $A \ldots C \rightarrow C \ldots A$ satisfies
$qW_N(A......C)=W_N(C......A)$, as long as $N_A = N_B =N_C$. Note that
when the three densities are not equal, the system supports a current,
and thus it cannot satisfy detailed balance. \

In order to demonstrate phase separation it is sufficient to show that
for large finite $k$ the two point density correlation function satisfies
\begin{equation}
\label{eq:lim}
\lim_{N \to \infty}(\langle A_1A_k \rangle - \langle A_1 \rangle
\langle A_k \rangle) > 0\; .
\end{equation}
 In our case this implies
 $\lim_{N \to \infty}\langle
A_1A_k \rangle > 1/9$ for large $k$.  In fact we  show below that
in the limit of large $N$ and for any finite $k$:
\begin{equation}
\label{eq:A1Ak}
\langle A_1A_k \rangle = 1/3 - O(1/N),
\end{equation}
demonstrating the existence of phase separation in the model. This
result indicates that the phase separation is complete, in the sense
that the probability of finding, say, a $B$ or a $C$ particle a large
distance inside the $A$ domain vanishes in the thermodynamic limit.

We proceed by first calculating the partition sum, $Z_N$, showing that
for $q<1$ and to leading order in $N$
\begin{mathletters}
\begin{equation}
\label{eq:Z2}
Z_N  =  N/\left[ (q)_{\infty}\right] ^3 ,
\end{equation}
where
\begin{equation}
(q)_{\infty}  =  \lim_{n \rightarrow \infty} (1-q)(1-q^2) \ldots
(1-q^n). 
\end{equation}
\label{eq:Z2both}
\end{mathletters}
The partition sum for $q>1$ is obtained by replacing $q$ by $1/q$ in
this expression.  Note that the partition sum (\ref{eq:Z2both}) is
proportional to $N$ rather than being exponential in $N$. This is a
consequence of the long range interactions; low energy excitations are
localized about the domain boundaries and their degeneracy does not
increase with system size (in contrast to short range models).  At
$q=1$, however, all configurations are equally probable and the
partition sum is exponential in $N$.

The calculation of the partition sum is greatly simplified by noting
that configurations with energy larger than $aN$, where $a>0$ is a
constant, can be neglected in the thermodynamic limit.  Here we show
that this is the case for $q<(1/3)^{1/a}$. A more detailed analysis
completes the proof for all $q \neq 1$ \cite{our}.  Consider the sum
$Z_{m > a N}=\sum_{m=a N+1}^{N^2/9}D(m)q^{m}$, where $D(m)$ is the
number of states of energy $m$. Noting that $D(m)$ is bounded from
above by the total number states which in turn is bounded by $3^N$,
one finds that $Z_{m > a N}$ decays exponentially with $N$ as long as
$q<(1/3)^{1/a}$.  Hence, the partition sum may be replaced by the
truncated one as long as exponentially small corrections in $N$ are of
no interest.

To calculate the truncated partition sum we note that states with
energy $m< N/3-1$, may be uniquely decomposed into $N$ disjoint sets,
where each set corresponds to one of the $N$ ground states of the
system.  Any state within a given set may be obtained from the
corresponding ground state by successive permutations of nearest
neighbor particles, in a way that the energy always increases along
the path of intermediate states.

The sets are labeled by $l=1 \ldots N$, the position of the rightmost
$A$ particle in the $A$ domain in the ground state configuration of
that set (Fig.(1)). Due to the translational invariance of the model
the partition sum can be written as $Z_N= N {\cal Z} + e^{-O(N)}$
where ${\cal Z}$ is the partition sum obtained by summing over one of
the $N$ sets of configurations.
\begin{figure}
\center{\psfig{figure=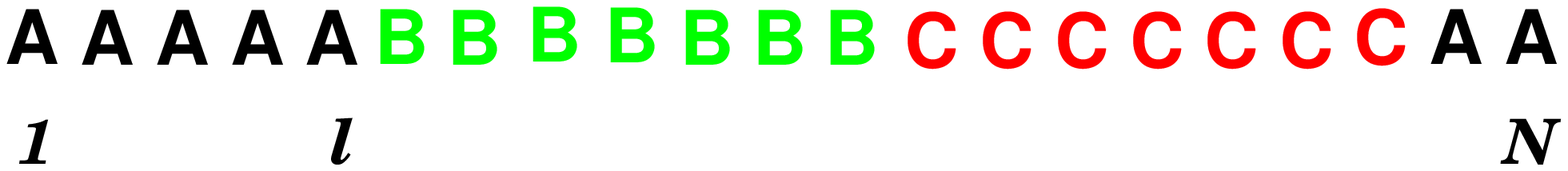,width=8cm}}
\caption{The $l=5$ ground state for an  $N=21$ system.}
\end{figure}
We now turn to calculating ${\cal Z}$. Consider an excitation of
energy $m$ created only at one domain boundary, say $AB$.  The
excitation can be formed by one or more $B$ particles moving into the
$A$ domain (equivalently $A$ particles moving into the $B$ domain). A
moving $B$ particle may be considered as a walker. The energy of the
system increases linearly with the distance traveled by the walker
inside the $A$ domain. An excitation of energy $m$ at the $AB$
boundary is formed by $j$ walkers passing a total distance of
$m$. Hence, the total number of states of energy $m$ at the $AB$
boundary is equal to the number of ways $P(m)$ of partitioning an
integer $m$ into a sum of (positive) integers. When considering all
three interfaces in the ground state, the number of states composed of
excitations of energy $m_i , i=1,2,3$ at the three interfaces is given
by $P(m_1) P(m_2) P(m_3)$. Note that since only excitations with total
energy smaller than $N/3-1$ are considered, a walker can not travel
from one boundary to another, and local excitations at the three
boundaries are independent.  The partition sum ${\cal Z}$, after
taking the thermodynamic limit, is given by
\begin{equation}
\label{eq:Z}
{\cal Z}= \sum_{m = 0}^{\infty}q^m \sum_{m_i =
0}^{m}\! P(m_1) P(m_2) P(m_3) \delta_{m_1 + m_2 +m_3,m}.
\end{equation}
 This sum may be rewritten as
\begin{equation}
\label{eq:Z10} 
{\cal Z}= (\sum_{m =0}^{\infty} q^{m} P(m))^3.
\end{equation}
Using the classical result for the generating function of $P(m)$
\cite{andrew}, attributed to Euler, we obtain
Eqs.(\ref{eq:Z2both}). Note that excitations of energy up to $N/3-2$
are properly accounted for. Once excitations of greater energy are
considered, triplets of the form $\ldots ABC \ldots$ can be
formed. These triplets can be moved around the lattice with no energy
cost so that the degeneracy is not accounted for correctly in
Eqs.(\ref{eq:Z2both}).  However, our results above imply that these
contributions are unimportant for $q < (1/3)^3$.  A more detailed
analysis \cite{our} confirms that for all $q< 1$,
Eqs. (\ref{eq:Z2both}) give the partition sum up to corrections
exponentially small in $N$.

In order to calculate the correlation function considered in the next
paragraphs, it is useful to introduce a partial sum ${\cal Z}_s$. It
is defined as the partition sum under the constraint that one of the
walkers at the $AB$ interface has traveled a distance $s$. One can
show that
\begin{equation}
\label{eq:Zs}
{\cal Z}_s =q^s{\cal Z}.
\end{equation}
We now turn to the correlation function $\langle A_1A_k\rangle
=1/3-\langle A_1B_k \rangle - \langle A_1C_k \rangle $. In order to
prove Eq. (\ref{eq:A1Ak}) we show that the correlation functions
$\langle A_1B_k \rangle$ and $\langle A_1C_k\rangle$ are of
$O(1/N)$. This is done explicitly for $\langle A_1B_k
\rangle$. Similar considerations yield the same result for $\langle
A_1C_k\rangle$. Define $\langle A_1B_k \rangle_l$ as the correlation
function calculated within the set of states related to the ground
state $l$.  Neglecting corrections exponentially small in $N$ we can
write
\begin{equation}
\label{eq:ineql}
\langle A_1B_k \rangle = \frac{1}{N} \sum_{l=1}^{N}\langle
A_1B_k\rangle_l.
\end{equation}

We now outline the proof that $\sum_{l=1}^{N}\langle A_1B_k\rangle_l$
is finite so that for large $N$, $\langle A_1B_k \rangle $ vanishes
like $1/N$. Consider first the $l=1,...,N/3$ terms in the sum
(\ref{eq:ineql}) for which $A_1 =1$ in the ground state.  The sum of
these terms may be broken into two parts in the following way:
\begin{equation}
\label{eq:case1}
\sum_{l=1}^{N/3}\langle A_1B_k\rangle_l = \sum_{l=1}^{k-1}\langle
A_1B_k\rangle_l + \sum_{l=k}^{N/3}\langle A_1B_k\rangle_l.
\end{equation}
To bound this sum we note that the first term contains contributions
from states corresponding to ground states in which $A_1 = B_k
=1$. Here we use the bound $\langle A_1B_k\rangle_l \leq 1$ for
$l=1,...k-1$. The second term contains contributions from states
corresponding to ground states in which $A_1=1$ but $B_k = 0$. Here,
only excited states in which one of the B walkers travels at least a
distance $l-k+1$ into the $A$ domain may contribute to $\langle
A_1B_k\rangle_l$.  Thus, for $k \leq l \leq N/3$ the correlation
function $\langle A_1B_k\rangle_l$ may be bounded from above by
$\sum_{s=l-k+1}^{\infty}{\cal Z}_{s}/{\cal Z}$. Combining the bounds
for the two parts one finds
\begin{equation}
\label{eq:case1pt2}
\sum_{l=1}^{N/3}\langle A_1B_k\rangle_l \leq k-1 +
\sum_{l=k }^{N/3} \sum_{s=l-k+1}^{\infty}q^s\; .
\end{equation}
The geometrical series in (\ref{eq:case1pt2}) converges for $q<1$ and
thus the sum $\sum_{l=1}^{N/3}\langle A_1B_k\rangle_l$ is bounded in
the thermodynamic limit. Similar considerations yield a bound for the
remaining $2N/3$ terms in (\ref{eq:ineql}) demonstrating that $\langle
A_1B_k \rangle =O(1/N) $.  We have thus shown that $\langle A_1A_k
\rangle = 1/3 - O(1/N)$ for large $N$ and any finite $k$,
demonstrating phase separation in the model.\

So far we have presented some results and bounds in the thermodynamic
limit. It is also of interest to obtain exact results for finite rings
to investigate how the limit is approached.  We do this by applying a
matrix ansatz method which has recently been introduced for studying
$1d$ non-equilibrium systems \cite{Der}.  Generalizing this approach
to replace the matrix product used as steady state ansatz by a tensor
product we found that we could apply the method to the model
(\ref{eq:dynamics}) for the case $N_A=N_B=N_C$. Details of these
calculations will be presented elsewhere \cite{our}, however we would
like to present some exact results for small systems obtained by this
method. We have calculated the correlation function $\langle A_1
A_{N/2} \rangle$ which provides a measure of phase separation. In a
disordered state this correlation function is equal to $(N-3)/9(N-1)$
and approaches $1/9$ in the large $N$ limit. It should be smaller for
a phase separated state. In fact we find it approaches zero over a
range of $q$ values which increases with $N$ (Fig.(2), inset).  To
investigate the finite size scaling near the $q=1$ (infinite
temperature) critical point it seems natural to choose $N \ln q$ as a
scaling variable.  This variable represents the ratio of domain wall
width ($ 1/\vert \ln q\vert $) to domain size ($N/3$).  In Fig.(2) the
scaling collapse for small systems is illustrated.
\begin{figure}
\psfig{figure=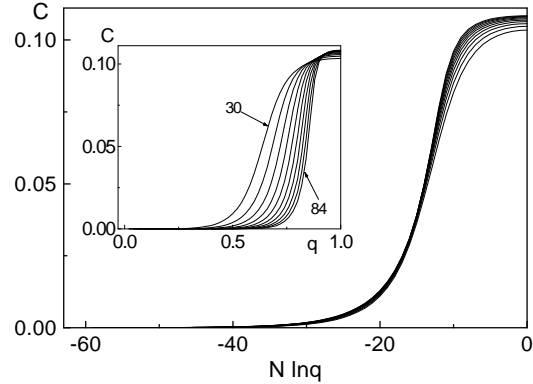,width=8cm}
\caption{ The correlation function $C=\langle A_1 A_{N/2} \rangle $ as
a function of the scaled variable $N \protect{\ln} q$ for $N=
30,36,42,\ldots,84$.  The inset shows the same data plotted against
$q$. Note that at $q=1$ the curves approach $1/9$ as $N$ increases
(see text). }
\end{figure}

{\it C. Discussion:}
The analysis presented above dealt with the case $N_A=N_B=N_C$. In the
general case where the densities of the three species are not equal,
detailed balance is not satisfied.  However, the heuristic arguments
for phase separation given at the beginning of the Letter are expected
to hold for the general case provided that none of the densities is
zero. Namely, configurations of the type $ \ldots AAABBBBCCAAA \ldots$
are stable, and the time for a totally phase separated state to break
up grows exponential in $N$.  Numerical simulations of the model
supports the existence of phase separation \cite{our}.

We thank D. Kandel, J. L. Lebowitz, S. Ramaswamy and E. R. Speer for
interesting discussions.  The support of Minerva Foundation, Munich,
Germany (DM), The Royal Society and The Einstein Center (MRE) are
gratefully acknowledged.


\end{document}